\documentstyle[12pt]{article}

\begin{document}
\sloppy
\begin{flushright}{KEK-TH-517\\ May '97}\end{flushright}
\vskip 1.5 truecm
\centerline{\large{\bf Successful D-term Inflation }}
\centerline{\large{\bf with moduli}}
\vskip .75 truecm
\centerline{\bf Tomohiro Matsuda
\footnote{matsuda@theory.kek.jp}}
\vskip .4 truecm
\centerline {\it National Laboratory For High Energy Physics (KEK)}
\centerline {\it Tsukuba, Ibaraki 305, Japan}
\vskip 1. truecm
\makeatletter
\@addtoreset{equation}{section}
\def\theequation{\thesection.\arabic{equation}}
\makeatother
\vskip 1. truecm
\begin{abstract}
\hspace*{\parindent}
We  examine a natural extension of D-term inflation and construct a 
successful model. General type of the D-term  potential is shown to 
produce  successful inflation with appropriate COBE normalization.
The effect of the dilaton shift at the inflation period is  taken 
into account.
\end{abstract}
\newpage
\section{Introduction}
\hspace*{\parindent}
At present, supersymmetry seems to provide the most likely solution to the
problem of the large hierarchy between weak and GUT or Planck
scales.
As there is yet no direct experimental evidence for supersymmetry,
it is worth turning to the early Universe for possible signatures.
Many models of supersymmetry breaking involve many flat directions,
which means that there exist many particles with
weak scale mass and Planck scale suppressed couplings.
Coherent production of such particles in the early Universe destroys
the successful prediction of nucleosynthesis.
As is discussed in ref.\cite{weak}, this problem may be solved by a
brief period of weak scale inflation.
This weak scale inflation has already shown to be realized by thermal
\cite{thermal} or parametric resonance\cite{pr} models.
On the other hand, in general supergravity theories, 
chaotic inflation suffers from some difficulties.
The main reason is that the minimal supergravity potential has
an exponential factor which prevents natural realization of 
chaotic inflation.
For  example, let us consider  superpotential of this form:
\begin{equation}
  W=\frac{\lambda}{3}\Phi^{3}
\end{equation}
where $\Phi$ is a gauge singlet superfield in the hidden sector
of the theory.
The scalar potential for the inflation field is obtained from 
superpotential as:
\begin{equation}
  V\simeq e^{\frac{\phi^{*}\phi}{M_{p}^{2}}}|\lambda\phi^{2}|^{2}
  + higher terms
\end{equation}
where $\phi$ represents the scalar component of $\Phi$.
Demanding that the e-foldings are larger than 60 ($N_{e}>60$), 
the resulting constraint is 
\begin{eqnarray}
  N_{e}&\simeq&\frac{8\pi}{M_{p}^{2}}\int^{\phi_{0}}_{M_{p}}
  \left(\frac{V}{V'}\right)d\phi\nonumber\\
    &=&\frac{\pi}{M_{p}^{2}}[\phi^{2}_{0}-M_{p}^{2}]>60
\end{eqnarray}
where we have assumed that chaotic inflation ends at 
$\phi\sim M_{p}$.
This means that the initial value of $\phi$ should be larger 
than $4\sim 5 M_{p}$.
The upper bound for $\lambda$ is given by the constraint on the Hubble 
parameter which results from the bound for density fluctuations.
\begin{equation}
  H\le10^{14}GeV
\end{equation}
As a result, the upper bound for $\lambda$ is $e^{16}\lambda^{2}
\le 10^{-13}$ i.e. $\lambda^{2}<10^{-22}$.
This bound is about $10^{-10}$ times smaller than 
the one for ordinary chaotic inflation.
In addition to this fine-tuning problem, slow-roll condition is
also problematic.
To avoid these difficulties, we can choose a potential as:
\begin{equation}
W=m\Phi^{2}  
\end{equation}
Combined with cosmic strings\cite{string}, 
hybrid inflation\cite{hybrid}
or non-trivial K\"ahler potential\cite{yanagida},
we can make successful scenarios.
But here we do not mention these another possibilities.
Recently, it was proposed that a variant of hybrid inflation
combined with a D-term potential can solve these difficulties
\cite{Dinfla}.
(We can also find the idea of D-term inflation
in \cite{add}.)
For example, let us consider an anomalous $U(1)$ gauge group
which is motivated from superstring effective theories\cite{anod}.
The D-term potential for the anomalous $U(1)$ is then
\begin{eqnarray}
V_{0}&=&\frac{g^{2}}{2}
\left||\phi_{+}|^{2}-|\phi_{-}|^{2}+M^{2}\right|^{2}\nonumber\\
M^{2}&=&\frac{TrQ_{A}}{192\pi^{2}}g^{2}M_{p}^{2}.
\end{eqnarray}
Here $g$ is the coupling constant for the anomalous $U(1)$ gauge
group, $\phi_{\pm}$ is a scalar field which has charge $\pm1$ under the
anomalous $U(1)$ gauge group and $TrQ_{A}$ is the sum of the charge
of the fields.
We can calculate $M$ in some specific models\cite{anod}
and is estimated to be $10^{-1}\sim 10^{-2}$ smaller than Planck mass.
In addition to the  scalar potential that is motivated from
the anomalous $U(1)$ D-term, we should
also include  superpotential(F-term) which contains a coupling
between $\phi_{\pm}$ and a singlet $\sigma$:
\begin{equation}
W=\lambda\sigma\phi_{+}\phi_{-}
\end{equation}
and  K\"ahler potential
\begin{eqnarray}
K&=&\sigma\overline{\sigma}+\phi_{\pm}\overline{\phi}_{\pm}.
\end{eqnarray}
The explicit form of the total scalar potential in the global 
supersymmetry limit is then
\begin{eqnarray}
V&=&|\lambda\sigma|^{2}
(|\phi_{+}|^{2}+|\phi_{-}|^{2})+|\lambda\phi_{+}\phi_{-}|^{2}
\nonumber\\
&&+\frac{g^{2}}{2}(|\phi_{+}|^{2}-|\phi_{-}|^{2}+M^{2})^{2}.
\end{eqnarray}
This potential has two types of  minima, one at $\phi_{\pm}=0$ and $\sigma$ 
large but undetermined,
and the other at $|\phi_{-}|^{2}=M^{2}$ and $|\phi_{+}|=\sigma=0$.
The former is not a true minimum.
Hybrid inflation occurs for large $\sigma$ and small $\phi_{\pm}$
when the vacuum energy is dominated by the D-term:
\begin{equation}
V_{D}=\frac{g^{2}}{2}M^{4}\equiv 3H^{2}M_{p}^{2}.
\end{equation}
Here, Hubble constant $H^{2}$ is given by $H^{2}
\sim g^{2}M^{4}/M_{p}^{2}$.
In this region, the mass for $\phi_{\pm}$ is estimated to be:
\begin{equation}
m^{2}_{\phi_{\pm}}=|\lambda|^{2}\sigma^{2}\pm g^{2}M^{2}
\end{equation}
For large $\sigma$, this mass term becomes large and it drives $\phi_{\pm}$
to its minimum at $\phi_{\pm}=0$ very quickly.
After $\phi_{\pm}$ settled down to $\phi_{\pm}=0$, 
the mass for $\sigma$ vanishes
at the tree level, but is lifted by 1-loop correction which is induced
by a mass splitting in the $\phi_{\pm}$ superpartners.
The explicit form of 1-loop potential along the flat direction
which is parametrized by $\sigma$ is 
\begin{equation}
V_{1-loop}=\frac{g^{2}}{2}M^{4}\left(1+\frac{1}{16\pi^{2}}log
\frac{\lambda^{2}\sigma^{2}}
{\Lambda^{2}}\right).
\end{equation}
Here $\Lambda$ is the renormalization scale.
The mass for $\sigma$ is thus written as:
\begin{equation}
m_{\sigma}^{2}=\frac{g^{2}M^{4}}{16\pi^{2}\sigma^{2}}.
\end{equation}
Inflation occurs for $|\sigma|\ge\sigma_{c}=gM/\lambda$.
At the end of inflation
$m_{\phi_{-}}$ becomes negative and then $\phi_{-}$ begins to roll down
the potential till it reaches to its true minimum at $\phi_{-}=M$.
The resulting density fluctuation is\cite{hybrid} 
\begin{equation}
\frac{\delta \rho}{\rho}\sim\frac{\lambda 
g^{2}M^{5}}{M_{p}^{3}m_{\sigma}^{2}}.
\end{equation}
To obtain a correct order of the density fluctuation
$\delta \rho/\rho \sim 10^{-5}$, we should make
some unnatural assumptions.
We should set $M\sim M_{P}/300$ which is about $10^{-2}$ times smaller
than the one obtained from a naive consideration.
In general one sets $TrQ_{A}\sim 100$ and $g^{2}\sim 1/4$ then obtains 
$M^{2}\sim M_{p}^{2}\times 10^{-2}$, i.e. $M\sim M_{p} \times 10^{-1}$.
Even if we set $TrQ_{A}=1$, we obtain at least $M \sim M_{p}\times 10^{-2}$.
We should also note that whenever we deal with a theory of stringy nature,
we should  inevitably   consider the effects of  moduli fields.
For example, let us consider the dilaton dependence of the gauge
coupling constant $g$.
The simplest form of $g$ is 
\begin{equation}
\frac{1}{g^{2}}\equiv f(S)=kS
\end{equation}
where $k$ is an integer  and $S$ is a dilaton superfield.
In the following we set $k=1$ for simplicity.
This dependence is very crucial for the above observation because
the D-term potential is now depends on the inverse of $S$,
thus it pushes  $S$  away to the infinity.
The explicit form of the scalar potential is 
\begin{equation}
V_{D}=\frac{1}{2S} M^{4}.
\end{equation}
On the other hand, when we consider an inflation scenario 
with supersymmetry we should also solve the Polonyi problem which 
comes from the flat directions or almost flat directions in
supersymmetric theories.
These flat directions or moduli are very common in supersymmetric
theories especially in superstring motivated ones.
A few years ago, L.Randall and S.Thomas\cite{weak} have shown that
the  second weak inflation can solve this problem and in the subsequent
papers it was discussed that thermal inflation\cite{thermal}
or non-thermal fluctuation induced by  parametric resonance\cite{pr}
can be the candidates for this type of weak inflation.
We can also find an observation of the Polonyi problem for dilaton
and moduli in \cite{casas}.
In this paper we propose a successful scenario for D-term
inflation with a dilaton dependent gauge coupling.
We also construct a successful chaotic
inflation scenario with $\phi^{4}$ type potential in supersymmetric 
theories and 
show that a general type of an anomalous $U(1)$ D-term 
can naturally induce chaotic  inflation.
In this model the shift of dilaton at the inflation period plays 
a crucial role.
Of course considering such effects of the 
moduli shift at the inflation period
is very important when we consider a string motivated supergravity theory.
\section{Successful inflation with a D-term potential}
\hspace*{\parindent} 
In this section we consider the case in which $g$ in the inflation period
is dynamically determined.
Here  we do not assume that $g$ is always the
same value throughout the early stage of the Universe.
First, let us consider a naive extension of the above scenario
for D-term inflation.
As we have noted in the previous section, 
dilaton field $S$ is always pushed away to larger value 
at the early stage of  inflation.
The explicit form of the potential at inflation period is 
\begin{eqnarray}
\label{vd1}
V_{D}&=&\frac{M^{4}}{2S}.
\end{eqnarray}
If the potential for $S$ is not stabilized, $S$ remains large
and we can never reach at the real world.
Of course, this is a  common situation for the string
motivated supergravity models and we know that we cannot construct
phenomenologically viable models unless we assume that 
the dilaton is really stabilized by some (unknown) mechanisms.
The expected form of the  dilaton potential is written in
many variants.
The simplest form will be 
\begin{equation}
V_{S}=m^{4}S^{2}
\end{equation}
or
\begin{equation}
V_{S}=m^{2}M_{p}^{2}S^{2}
\end{equation}
where m is some
intermediate scale $m_{w}\le m \le M_{p}$.
We neglect the imaginary part of the dilaton superfield and simply
assume that $S\sim ReS$ because the imaginary part is not important 
in the following discussions.
One may expect many other terms  suppressed by $n$ powers of Planck scale,
but here we do not consider such complicated extensions of this scenario
but instead we consider only one such example:
\begin{equation}
V_{S}=\frac{[m^{3}]^{2}}{M_{p}^{2}}S^{2}.
\end{equation}
For our purpose the potential for large $S$ 
can be characterized simply
by its power of the dilaton field $S$, thus here we do not write down
the explicit form of the potential.
It seems very natural to expect that the dilaton potential is
stabilized by some mechanisms, but the precise form of the potential
is  still unknown.       
We may also expect that the dilaton potential is derived from the M-theory
motivated models. 
One instance is recently derived in ref.\cite{Mtheory} and shown to take
the form:
\begin{eqnarray}
V_{S}\sim m^{4}\frac{|1-S|^{2}}{S+\overline{S}}.
\end{eqnarray}
This potential behaves like $V_{S}\sim m^{4}S$ for large $S$.
We use these  
potentials and see what happens to the density fluctuation.
The total scalar potential for $S$ is now obtained:
\begin{eqnarray}
V_{tot}&=&V_{D}+V_{S}.
\end{eqnarray}
At the early stage of the Universe, when $V_{D}$ is given by (\ref{vd1}),
$S$ takes the following value
for each possible potentials.
\begin{eqnarray}
V_{S}=m^{4}S^{2}&;& S \sim \left(\frac{M}{m}\right)^{4/3}\\
V_{S}=\frac{[m^{3}]^{2}}{M_{p}^{2}}S^{2} &;& 
S\sim\left(\frac{M^{4}M_{p}^{2}}{m^{6}}\right)^{1/3}\\
V_{S}=m^{4}S &;& 
S\sim\left(\frac{M}{m}\right)^{2}
\end{eqnarray}
In any case, $S$ becomes large in the inflation phase 
and it makes the density fluctuation
smaller than the one derived with an assumption
that $S$ should be fixed at $S\sim 1$ throughout the inflation.
Let us assume  $TrQ_{A}\sim 10^{2}$ and $\lambda\sim 1$ then the constraint 
for the density fluctuation is satisfied for $S\sim 10^{1\sim 2}$.
This requirement is satisfied when the intermediate scale $m$ is comparable 
to Planck mass $M_{p}$, which means that dilaton is
stabilized at very high energy scale like the string scale.
The assumption that dilaton  is stabilized at high energy
scale is favorable from phenomenological point of view and sometimes
it is utilized without mentioning the specific mechanism.
However, as we have shown above, the shift of the dilaton field at inflation
is important even if we assume that the dilaton potential is stabilized
at  high energy scale. 
For the second example, let us consider an alternative of D-term
inflation and show that small $m$ is also available for successful
inflation.
Here we consider  an anomalous U(1) gauge group again.
The D-term potential and  superpotential are the same as
before.
The total scalar potential is
\begin{eqnarray}
\label{these1}
V&=&|\lambda\sigma|^{2}
(|\phi_{+}|^{2}+|\phi_{-}|^{2})+|\lambda\phi_{+}\phi_{-}|^{2}
\nonumber\\
&&+\frac{g^{2}}{2}(|\phi_{+}|^{2}-|\phi_{-}|^{2}+M^{2})^{2}\nonumber\\
&&+V_{S}.
\end{eqnarray}
Here we consider a case when chaotic inflation starts with large
$\phi_{+}$ or $\phi_{-}$.
This is possible when $g$ is small(i.e. $S$ is pushed away to very 
large value)
and $\phi_{+}(\phi_{-})$ is  $O(M_{P})$ before inflation.
Unlike ordinary supersymmetric chaotic inflation scenarios, 
this potential does not have any exponential factor that makes it 
difficult to incorporate chaotic inflation with $\phi^{4}$
type potential.
The upper bound for the coupling constant $g^{2}$ is the same as
ordinary chaotic inflation  and 
is required to be $g^{2}\sim O(10^{-13})$ i.e.$S\sim 10^{13}$.
In this case, $V_{D}$ is
\begin{eqnarray}
\label{VD}
V_{D}&\sim&\frac{(5M_{p})^{4}}{2S}.
\end{eqnarray}
The required value for $S$ ($S\sim 10^{13}$)
is obtained if $m$ takes the following value
for each possible dilaton potential.
\begin{eqnarray}
V_{S}=m^{4}S^{2}&;& m\sim 10^{8}Gev\\
V_{S}=\frac{[m^{3}]^{2}}{M_{p}^{2}} S^{2}&;& 
m\sim 10^{11}Gev\\
V_{S}=m^{4}S &;& 
m\sim 10^{11}Gev
\end{eqnarray}
We can relate these scales to the  supersymmetry breaking scale 
or the scale of the fifth dimension\cite{Mtheory}. 
Let us see whether the second weak inflation can be realized
in these models.
Generally, the second weak inflation is expected to be realized by
some flat directions thus
we focus attention to the question 
whether the second stage of inflation can be realized in some flat
directions in the theory.
We know that the second stage of   thermal inflation
can take place in common supersymmetric models\cite{thermal}
without any additional potential, but sometimes it is insufficient.
What we want to see here is whether a minimal set of the D-term potential
given above can help weak scale inflation.
For this purpose, let us consider  weak scale inflation driven 
by  parametric resonance\cite{pr}.
The first model described above is not 
suitable for this purpose, thus we consider only the second one.
Here we consider a case where some flat directions in the theory
couples to $\sigma$ through
\begin{eqnarray}
\label{flat}
  V(X_{F},\sigma)&=&-\frac{m^{2}_{soft}}{2}X_{F}^{2}
+\lambda'^{2}X_{F}^{2}\sigma^{2}
+higher terms.
\end{eqnarray}
where $X_{F}$ represents some flat directions and $m_{soft}$ is the
soft breaking mass.
According to \cite{pr}, these potentials (\ref{these1}) and (\ref{flat})
can produce an additional period 
of non-thermal inflation.
The Universe expands in the ratio
\begin{equation}
  \frac{a_{c}}{a_{0}}\sim \left(\frac{\lambda'^{2}}{g^{2}}\right)^{1/4}.
\end{equation}
In our case, we obtain $a_{c}/a_{0}\sim 10^{3}$
for $\lambda'\sim O(1)$
which is enough to solve the Polonyi problem.
However, a serious problem arises when we consider the oscillation
of the dilaton.
As a result, the  same problem will be there for the dilaton
which  should be solved again by thermal inflation.
However, it seems very interesting that the problem is reduced
to the displaced dilaton which sometimes plays specific roles.
We hope that the mechanism stated above helps to solving the 
Polonyi problem, but for now a further stage of thermal 
inflation is still required. 
\section{Conclusion}
\hspace*{\parindent}
We have examined a natural extension of D-term inflation
and constructed a successful model.
The effect of the dilaton shift at the inflation period
is taken into account.
It is also important to note that the model we have constructed
requires no fine tuning of parameters to realize chaotic inflation.
\section{Acknowledgment}
We thank  K.Fujikawa and K.tobe  for useful discussions.

\end{document}